\documentclass[12pt]{iopart}

\usepackage{iopams}
\usepackage{graphicx}

\begin{document}

\title[On the fidelity of quantum information processing with Rydberg atoms]%
{On the fidelity of quantum information processing with Rydberg atoms:
Role of the non-molecular resonances}

\author{Yurii V Dumin$^{1,2}$}

\address{$^1$ Max Planck Institute
for the Physics of Complex Systems (MPIPKS),\\
Noethnitzer Str.\ 38, 01187 Dresden, Germany}

\address{$^2$ Sternberg Institute (GAISh) of
Lomonosov Moscow State University,\\
Universitetski prosp.\ 13, 119992 Moscow, Russia}

\eads{\mailto{dumin@pks.mpg.de}, \mailto{dumin@yahoo.com}}

\begin{abstract}
Rydberg blockade of ultracold atoms is considered now as one of
the most promising tools for the implementation of quantum computing,
but its fidelity can be substantially compromised by detrimental excitation
of the neighbouring atoms.
This phenomenon has been investigated recently in detail for the particular
case of molecular resonances (i.e., resulting in the formation of quasi-bound
states).
However, as will be shown in the present paper, an even greater effect
can come from the non-molecular resonances, which therefore should be
taken into account very carefully.
\end{abstract}

%

\submitto{\jpb}

\maketitle

\section{Introduction}
\label{sec:Intro}

The phenomenon of Rydberg blockade~--~impossibility of simultaneous excitation
of the nearby Rydberg atoms due to the shifts of their energy levels~--~was
suggested as a tool for quantum information processing by Lukin, et al.
in the early 2000's~\cite{Lukin_01}.
A few years later, the practical feasibility of such a protocol was confirmed
experimentally~\cite{Tong_04,Singer_04}, and now it is widely discussed in
the context of quantum computing~\cite{Saffman_10}.

Unfortunately, as was qualitatively discussed in our paper~\cite{Dumin_14},
an important problem that can arise in this way is the possibility of
detrimental excitation of a nearby atom (within the standard blockade radius)
caused by the strongly-perturbed energy levels with neighbouring values of
the quantum numbers (see, for example, figure~1 in paper~\cite{Dumin_14}
as well as figure~\ref{fig:En_curves} below).
Later, this effect was studied in detail in the work~\cite{Derevianko_15}.
In particular, the calculation was performed both for the positions and
excitation rates of the major resonances forming the quasi-molecular states
of two Rb atoms that are asymptotically in $100s$ states, and
the corresponding consequences for the fidelity of the quantum computing
were discussed.

However, as will be shown below, the total number of resonances breaking
the Rydberg blockade (i.e., resulting in the simultaneous excitation of
two nearby Rydberg atoms) should be much greater than in figure~1 of
the above-cited paper.
The most of these resonances are non-molecular, i.e., do not necessarily
result in the formation of binding potentials.
Anyway, they can break the Rydberg blockade and, therefore, their detrimental
effect on the fidelity of quantum processing protocols must be kept in mind.

\section{Basic formulae}
\label{sec:Bas_form}

From our point of view, an efficient method for treating the Rydberg blockade
can be based on the idea of Stark splitting of the atomic energy levels by
the external electric field by a neighbouring already-excited atom.
This approach was pursued in our previous papers~\cite{Dumin_14,Dumin_15}
for the case of ``sequential'' excitation of two Rydberg atoms (i.e.,
at the time scales much greater than the inverse Rabi frequency), which is
actually most interesting for the experiments with ultracold Rydberg
plasmas~\cite{Robert_13}.
Here we are going to generalize this approach to the case of ``simultaneous''
excitation, which is immediately relevant to the quantum-computing
applications.
(Let us emphasize that the same simultaneous excitation was considered in
the work~\cite{Derevianko_15}.)

An important aspect of the Stark splitting that must be taken into account
in such analysis is the substantially nonuniform character of the dipolar
electric field produced by one Rydberg atom at the characteristic scale of
another atom.
As far as we know, the only special treatment of Stark effect in
the nonuniform field was done in 1970 by Bekenstein and
Krieger~\cite{Bekenstein_70}, who used a specific kind of the quasi-classical
approximation.
We performed such calculation beyond this approximation in
paper~\cite{Dumin_15} and found some corrections to the Bekenstein--Krieger
formula.
(However, these corrections are quite small for the low-angular-momentum
states, which are commonly employed in the experiments with Rydberg blockade.)

So, the general expression for Stark splitting of hydrogen-like energy
levels can be written in the atomic units as:
\begin{equation}
{\delta}E_n =
  E_n + \frac{ 1 }{ 2n^2 } =
  g_{1}{\cal E}_z - g_{2}{\cal E}_z^2
  + g_{3} \frac{ d {\cal E}_z }{ dz } \, ,
\label{eq:Stark_shift_gen}
\end{equation}
where the coefficients~$ g_{i} $ are
\begin{numparts}
\begin{eqnarray}
g_{1} \! & = &
  \frac{ 3 }{ 2 } \, n \Delta \, , \\
g_{2} \! & = &
  \frac{ n^4 }{ 16 }
  \left[ 17 n^2 - 3 {\Delta}^2  - 9 m^2 + 19 \right] , \\
g_{3} \! & = &
  \frac{ n^2 }{ 4 } \left[ 5 {\Delta}^2 + 2 n_1 n_2
  + ( n - m ) (m + 1) + 1 \right] .
\end{eqnarray}
\end{numparts}
\noindent
Here, $ \cal E $~is the electric field intensity,
$ n $~is the principal quantum number,
$ n_1 $ and $ n_2 $~are the parabolic quantum numbers
($ n_{1,2} \geqslant 0 $),
$ \Delta = n_{1} - n_{2} $~is the so-called electric quantum number, and
$ m $~is the \textit{absolute value} of the magnetic quantum number
(we accept here the Bethe--Salpeter designations~\cite{Bethe_57}).

As is known, the above-mentioned quantum numbers are related to each other as:
\begin{equation}
n = n_1 + n_2 + m + 1 \, ,
\end{equation}
so that the following inequalities are satisfied:
\begin{numparts}
\begin{eqnarray}
m \, \geqslant \, 0 \, , \\
n \, \geqslant \, m + 1 \, , \\
0 \, \leqslant \, n_1, n_2 \, \leqslant \, n - m - 1 \, .
\end{eqnarray}
\end{numparts}
Besides, it is easy to see that $ g_{2,3} \geqslant 0 $.

The first two terms in the right-hand side of
formula~(\ref{eq:Stark_shift_gen}) represent the well-known expressions
for the first- and second-order Stark effect in the uniform
field~\cite{Bethe_57,Gallagher_94,Landau_77},
and the third term takes into account the electric-field nonuniformity
(it was derived in our paper~\cite{Dumin_15} and, as have been already
mentioned above, slightly differs from the quasi-classical result by
Bekenstein and Krieger~\cite{Bekenstein_70}).
In principle, it might be possible to include here also the higher-order
corrections with respect to the electric field amplitude:
for example, the explicit expression with terms up to the fourth order
was obtained by Alliluev and Malkin~\cite{Alliluev_74},
and a general algorithm for deriving the terms of arbitrary order was
described by Silverstone~\cite{Silverstone_78}.
However, as follows from the subsequent analysis, such higher-order
corrections are completely negligible for our purposes%
\footnote{This fact is not surprising: we are interested in
the perturbation of energy levels on the order of difference between
the states with neighboring values of the principal quantum number;
while the higher-order corrections become significant on the scale of
the binding energy.
}.
However, the gradient term will be really important.

Next, let us consider two simultaneously-excited Rydberg atoms.
Each of them forms a dipolar electric field and, thereby,
disturbs its partner, producing the corresponding Stark splitting of
the energy levels.
So, we shall analyze further the mutually-perturbed diatomic system,
not assuming any specific kind of the interatomic interaction
(such as formation of the binding potential).

\begin{figure}
\center{\includegraphics[width=7.5cm]{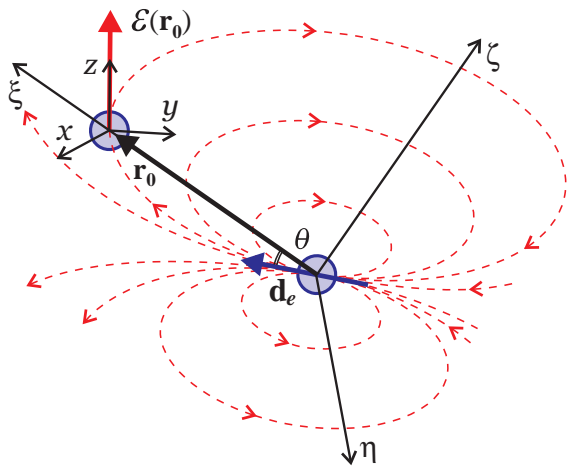}}
\caption{Sketch of the dipolar electric field produced by
the first atom at the position of the second atom.
\label{fig:Dip_field}}
\end{figure}

Let the first atom is located in the origin of the coordinate
system~($ \xi , \eta , \zeta $), and the position of the second atom
is given by the radius vector~$ {\bf r}_0 $ directed along $ \xi $-axis
(figure~\ref{fig:Dip_field}).
The electric dipole moment~$ {\bf d}_e $ of the first atom is tilted at
the angle~$ \theta $ with respect to this axis.
Next, the origin of the coordinate system~($ x , y , z $) is placed in
the location of the second atom, and its $ z $-axis is oriented along
the direction of the electric field in this point~$ {\cal E}({\bf r}_0) $;
so, by definition, the only nonzero component is~$ {\cal E}_z $.

The electric field potential is evidently given by formula:
\begin{equation}
\Phi = \frac{ {\bf d}_e \!\! \cdot \! {\bf r}_0 }{ r_0^3 } \, ;
\label{eq:Dip_poten}
\end{equation}
while the corresponding field intensity and its gradient can be written as
\begin{numparts}
\begin{eqnarray}
\; {\cal E}_z & = & \:
  \frac{ d_e }{ r_0^3 } \, ( 1 + 3 \, {\cos}^2 \theta )^{1/2} \, ,
\label{eq:Dip_field} \\[1ex]
\frac{ d {\cal E}_z }{ dz } & = &
  - \frac{ 3 d_e }{ r_0^4 } \:
  \frac{ 3 + 5 \, {\cos}^2 \theta }%
{ 1 + 3 \, {\cos}^2 \theta } \, \cos \theta \, .
\label{eq:Dip_field_grad}
\end{eqnarray}
\end{numparts}

For the sake of simplicity, we shall consider below in detail only the cases
of parallel and anti-parallel alignment of two dipoles.
Then, the above formulae are reduced to
\begin{equation}
{\cal E}_z = \frac{ 2 d_e }{ r_0^3 } \: ,
\quad
\frac{ d {\cal E}_z }{ dz } =
  -{\epsilon}_{\theta} \frac{ 6 d_e }{ r_0^4 } \: ,
\label{eq:Dip_field_par}
\end{equation}
where $ {\epsilon}_{\theta} = 1 $ at~$ \theta = 0 $, and $ -1 $
at $ \theta = {\pi} $.

Since the electric field~$ \cal E $ is considered here in the classical
approximation, then its source should be the expectation value of
the electric dipole operator $ \hat{\rm \bf d}_e = - \hat{\rm \bf r}_e $,
where $ {\rm \bf r}_e $~is the electron radius vector inside the atom.
The corresponding matrix element is well known, e.g., from calculations
of the first-order Stark effect~\cite{Bethe_57,Gallagher_94,Landau_77}.
So, for the first atom it will be equal to
\begin{equation}
d_e = - \langle \, \hat{\xi}_e \, \rangle =
  - \frac{3}{2} \, n \Delta \, .
\label{eq:Dip_ind}
\end{equation}
As can be easily seen,
\begin{equation}
{\epsilon}_{\theta} = -{\rm sign}({\Delta}) .
\label{eq:eps_theta_deriv}
\end{equation}

For subsequent applications it will be convenient to measure
the Stark energy shifts with respect to the energy of the unperturbed
state~$ \bar{n} $ whose blockade is studied and which will be denoted
by a bar:
\begin{equation}
\delta E_{\bar{n}} =
  \frac{ 1 }{ 2 {\bar{n}}^2 } - \frac{ 1 }{ 2 n^2 } + \delta E_n \, .
\label{eq:Ener_shift_redef}
\end{equation}
Besides, we shall normalize all lengths and energies to the characteristic
size and energy of the above-mentioned state~$ \bar{n} $, and the resulting
quantities will be marked by tildes:
\begin{equation}
r_0 = {\bar{n}}^2 \tilde{r} \, , \quad
E = \tilde{E} / ( 2 {\bar{n}}^2 ) \, .
\label{eq:Scaled_quant}
\end{equation}
(For brevity, the scaled radius vector of the atom is written without
subscript~`0'.)

Finally, combining the formulae~(\ref{eq:Stark_shift_gen}) and
(\ref{eq:Dip_field_par})--(\ref{eq:Scaled_quant}),
we get the shifts of energy levels in the second atom produced by
the first atom:
\begin{eqnarray}
\delta \tilde{E}_{\bar n}^{(2)} & = & \:
  1 - \frac{ \bar{n}^2 }{ n^{(2)2} } \,
  + \, 9 \Bigg[ \frac{1}{{\tilde r}^3} \,
  \frac{ n^{(1)} \, n^{(2)} \,
  | {\Delta}^{(1)} | \, {\Delta}^{(2)} }{ \bar{n}^4 }
\nonumber \\[1ex]
&& - \, \frac{ 2 }{ {\tilde r}^6 } \,
  \frac{ g_2^{(2)} n^{(1)2} {\Delta}^{(1)2} }{ \bar{n}^{10} }
  +  \frac{ 2 }{ {\tilde r}^4 } \,
  \frac{ g_3^{(2)} n^{(1)} {\Delta}^{(1)} }{ \bar{n}^6 } \Bigg] ,
\label{eq:del_E(2)_gen}
\end{eqnarray}
where superscripts in parentheses denote the number of the atom
(to avoid its confusion with exponents).
The same expression with interchanged superscripts will evidently
give the energy shifts in the first atom.

To avoid cumbersome computations, we shall consider in detail
only two ``symmetric'' types of excitation in this diatomic system:
\begin{eqnarray*}
& \mbox{\bf (a)} & \;
{| n_1 , n_2 , m \rangle}^{\!(1)}\, {| n_1 , n_2 , m \rangle}^{\!(2)} ,
\; \mbox{i.e.},
\\ && \;
n \equiv n^{(1)} = n^{(2)} , \quad
\Delta \equiv {\Delta}^{(1)} = {\Delta}^{(2)} ,
\\
& \mbox{\bf (b)} & \;
{| n_1 , n_2 , m \rangle}^{\!(1)}\, {| n_2 , n_1 , m \rangle}^{\!(2)} ,
\; \mbox{i.e.},
\\ && \;
n \equiv n^{(1)} = n^{(2)} , \quad
\Delta \equiv {\Delta}^{(1)} = - {\Delta}^{(2)} .
\end{eqnarray*}
In other words, the atoms are excited either exactly to the same states
or to the states with interchanged parabolic quantum numbers.
As can be easily shown, case~(a) corresponds to parallel orientation of
the dipoles, while case~(b) represents the anti-parallel orientation
(either towards or away from each other).

So, the energy shifts in both atoms under the above assumptions will be
the same and equal to
\begin{eqnarray}
\delta \tilde{E}_{\bar n} \equiv \,
\delta \tilde{E}_{\bar n}^{(1)} \! =
\delta \tilde{E}_{\bar n}^{(2)} \! & = & \:
  1 - \frac{ \bar{n}^2 }{ n^2 }
  + \, 9 \Bigg[ \frac{1}{{\tilde r}^3} \,
  \frac{ n^2 {\Delta}^2 }{ \bar{n}^4 } \, \epsilon \: {\rm sign} ( \Delta )
\nonumber \\[1ex]
&& - \, \frac{2}{{\tilde r}^6} \,
  \frac{ g_2 n^2 {\Delta}^2 }{ \bar{n}^{10} }
  +  \frac{2}{{\tilde r}^4} \,
  \frac{ g_3 n \Delta }{ \bar{n}^6 } \Bigg] ,
\label{eq:del_E_sym}
\end{eqnarray}
where $ {\epsilon} = 1 $ and~$ -1 $ for parallel and anti-parallel
orientation of the dipoles, respectively.
Note that the first and second terms in the square brackets result from
the first- and second-order Stark effect in the uniform field, while
the third term comes from the first-order perturbation by
the electric-field gradient.

\section{Results of calculations}
\label{sec:Res_calc}

\begin{figure}
\center{\includegraphics[width=13cm]{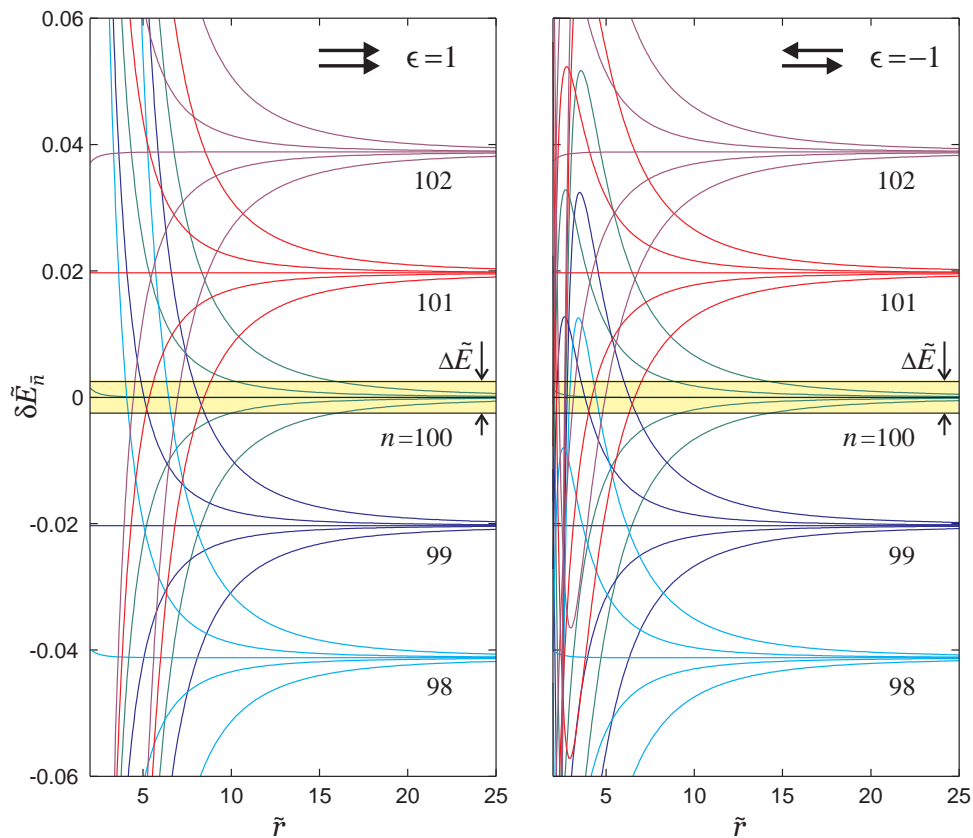}}
\caption{The Stark-split energy curves~$ \delta \tilde{E}_{\bar{n}} $
in each of the interacting atoms as function of distance~$ \tilde{r} $
between them for parallel ($ \epsilon = 1 $) and
anti-parallel ($ \epsilon = -1 $) orientation of the dipoles:
$ n = \! \bar{n} \! = 100, \: n_1 \! = \! 0, 25, 50, 75, 99 $ (green curves),
$ n \! = \! 99, \: n_1 \! = \! 0, 25, 49, 73, 98 $ (blue),
$ n \! = \! 98, \: n_1 \! = \! 0, 25, 49, 73, 97 $ (cyan),
$ n \! = \! 101, \: n_1 \! = \! 0, 25, 50, 75, 100 $ (red), and
$ n \! = \! 102, \: n_1 \! = \! 0, 25, 50, 75, 101 $ (violet).
The horizontal shaded (yellow) strip denotes the energy excitation
band of laser irradiation, $ \Delta \tilde{E} = 5{\cdot}10^{-3} $.
\label{fig:En_curves}}
\end{figure}

Now, the mathematical formalism outlined in the previous section can be
applied to the problem of Rydberg blockade.
Let us consider, for example, the blockade of $ 100s $~state of
the hydrogen-like atoms (i.e., $ \bar{n} = 100 $ and $ m = 0 $).
A sample of the Stark-split energy curves in the vicinity of this state
as function of the interatomic distance, calculated by
formula~(\ref{eq:del_E_sym}), are presented in figure~\ref{fig:En_curves}.
For better visibility, the excitation band of laser irradiation, shown by
a shaded strip along the horizontal axis, was taken here quite
large~$ \Delta \tilde{E} = 5{\cdot}10^{-3} $ (it is usually one or two
orders of magnitude less in the real experiments).

It is interesting to discuss briefly a significance of the various terms
in formula~(\ref{eq:del_E_sym}).
As follows from the detailed analysis, the dominant contribution comes from
the first-order uniform-field Stark effect.
The gradient term is always important and, moreover, qualitatively changes
the behavior of energy curves at small~$ \tilde{r} $ in the case of
anti-parallel dipoles.
The second-order uniform-field Stark effect is usually quite insignificant.
It becomes noticeable only at small distances in the case of anti-parallel
dipoles, when the first-order perturbations by the uniform field and
its gradient compensate each other to a large extent.

As can be seen in figure~\ref{fig:En_curves}, when two atoms approach each
other, the energy levels of the basic manifold~$ \bar{n} $, which have been
asymptotically degenerate and resonant with laser radiation at large
separations, experience an increasing perturbation and eventually leave
the bandwidth of the exciting irradiation.
As a result, the Rydberg blockade develops.
Nevertheless, when the interatomic separation decreases further,
the strongly-perturbed energy levels from the neighboring Stark manifolds
(with $ n \neq \bar{n} $) begin to enter the excitation band,
thereby restoring the possibility of excitation.
Hence, the Rydberg blockade should be broken at a set of
the ``resonant'' radii, where the energy curves intersect the horizontal
axis.
(The effect of avoided crossings of the energy levels was not plotted in
this figure just because we are interested only in the points of
intersection of the energetic curves with the horizontal axis, while
accurate identification of the corresponding states is of no importance
in this context.)

\begin{figure}
\center{\includegraphics[width=10cm]{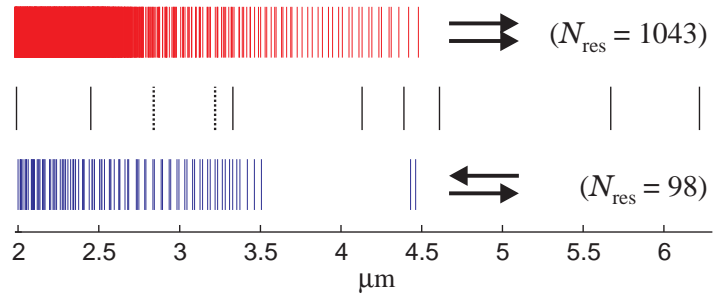}}
\caption{Diagram of the ``quasi-molecular'' resonances for $ 100s $~state of
Rb~\cite{Derevianko_15} (middle band) vs.\ all resonances for the same state
of hydrogen-like atoms (upper band for parallel orientation of the dipoles,
and lower band for the anti-parallel orientation).
Dotted bars denote the molecular resonances shown in
figure~1~\cite{Derevianko_15} but not listed in table~1 of the same paper.
\label{fig:Res_Diag}}
\end{figure}

The total number of the Stark-split manifolds participating in breaking
the Rydberg blockade cannot be specified exactly and is actually limited
by the scope of applicability of the perturbation theory.
This is roughly given by the condition that the energy shift should be
less than the absolute value of the unperturbed energy:
\begin{equation}
| E_n - E_{\bar{n}} | \lesssim | E_n | \, ,
\label{eq:Crit_perturb}
\end{equation}
implying that $ n \lesssim \sqrt{2} \: \bar{n} $ for the upper-lying
energy levels.
On the other hand, it can be easily seen that the perturbation theory
is applicable to all the low-lying levels, because
the criterion~(\ref{eq:Crit_perturb}) is always satisfied at $ n < \bar{n} $.

The most important fact following from our analysis is that the total number
of resonant radii turns out to be much greater than in the ``quasi-molecular''
approach~\cite{Derevianko_15}.
This is illustrated for the particular case of $ 100s $~states in
figure~\ref{fig:Res_Diag}: the number of resonances~$ N_{\rm res} $
at the distance~$ r_0 \geqslant 2\,{\mu}{\rm m} $ is an order of magnitude
greater than in the quasi-molecular approximation for the case of
anti-parallel orientation of the dipoles, and by two orders of magnitude
greater for the case of parallel orientation%
\footnote{
Let us emphasize that the patterns of the detrimental resonances are almost
insensitive to the bandwidth~$ \Delta \tilde{E} $ of the exciting irradiation.
Really, as can be seen in figure~\ref{fig:En_curves}, onset of the Rydberg
blockade strongly depends on~$ \Delta \tilde{E} $: the narrower is the width,
the greater is the radius of the blockade.
However, the radii of the blockade breaking are approximately insensitive
to~$ \Delta \tilde{E} $ because the corresponding energy curves intersect
the horizontal axis almost vertically.}.

\section{Conclusion}
\label{sec:Concl}

Unfortunately, it is impossible to compare the particular positions of
the resonances found in our calculations and in paper~\cite{Derevianko_15}
because of the considerable quantum defects in Rb, which was analyzed in
the last-cited work.
However, the striking disagreement apparent in figure~\ref{fig:Res_Diag}
can be hardly explained just by the quantum defects.
So, we should conclude that the quasi-molecular approach takes into account
only a small fraction of all the resonances breaking the Rydberg blockade.

In other words, formation of the bound (quasi-molecular) states is not
a necessary prerequisite for violating the Rydberg blockade.
The blockade will be broken each time when two nearby atoms are excited
simultaneously, irrelevant of the particular type of interaction between
them.
So, we believe that the detrimental influence of the non-molecular
resonances~-- which was overlooked before~-- should be taken into account
very carefully in any future applications of Rydberg blockade in
the quantum computing and other quantum-optics experiments.

\section*{References}


\end{document}